\documentclass[aps,prx,groupedaddress,reprint,showpacs,pdftex]{revtex4-1}

\usepackage{graphicx}
\usepackage{tabularx}
\usepackage{dcolumn}
\usepackage{bm}
\usepackage{amsmath}
\usepackage{amssymb}
\usepackage{txfonts}
\usepackage{url}
\usepackage{here}
\usepackage[usenames,dvipsnames]{color}
\definecolor{Gray}{gray}{0.0}
\definecolor{lightGray}{gray}{0.35}
\begin{document}
\title{
  \textit{ \textbf{ Ab initio }} search of polymer crystals with high thermal conductivity
}
\author{Keishu Utimula$^{1,*}$}
\author{Tom Ichibha$^{2}$}
\author{Ryo Maezono$^{2,3}$}
\author{Kenta Hongo$^{3,4,5}$}

\affiliation{$^{1}$
  School of Materials Science, JAIST, Asahidai 1-1, Nomi, Ishikawa,
  923-1292, Japan
}

\affiliation{$^{2}$
  School of Information Science, JAIST, Asahidai 1-1, Nomi, Ishikawa,
  923-1292, Japan
}

\affiliation{$^{3}$
  Computational Engineering Applications Unit, RIKEN, 
  2-1 Hirosawa, Wako, Saitama 351-0198, Japan
}

\affiliation{$^{3}$
  Research Center for Advanced Computing Infrastructure, JAIST,
  Asahidai 1-1, Nomi, Ishikawa 923-1292, Japan
}
\affiliation{$^{4}$
  Center for Materials Research by Information Integration,
  Research and Services Division of Materials Data and
  Integrated System, National Institute for Materials Science,
  Tsukuba 305-0047, Japan
}
\affiliation{$^{5}$
  PRESTO, Japan Science and Technology Agency, 4-1-8 Honcho,
  Kawaguchi-shi, Saitama 322-0012, Japan
}

\affiliation{$^{*}$
  mwkumk1702@icloud.com
}
\date{\today}
\begin{abstract}
Lattice thermal conductivities (LTC) for a subset of polymer crystals
from the Polymer Genome Library
were investigated to explore high LTC polymer systems.
We employed a first-principles approach to
evaluating phonon lifetimes
within the third-order perturbation theory
combined with density functional theory,
and then solved the linearized Boltzmann transport equation
with single-mode relaxation time approximated by the computed lifetime.
Typical high LTC polymer systems, polyethylene (PE) crystal and fiber, 
were benchmarked, which is reasonably consistent with previous references,
validating our approach.
We then applied it to not only typical polymer crystals, 
but also some selected ones having structural similarities to PE.
Among the latter crystals, 
we discovered that beta phase of Poly(vinylidenesurely fluoride) (PVF-$\beta$) crystal 
has higher LTC than PE at low temperature.
Our detailed mode analysis revealed that the phonon lifetime of PVF-$\beta$ is 
more locally distributed around lower frequency modes and four-times larger than that of PE.
It was also found from a simple data analysis that 
the LTC relatively correlates with curvature of energy-volume plot.
The curvature would be used as a descriptor for further exploration 
of high LTC polymer crystals by means of a data-driven approach beyond human-based one.
\end{abstract}
\maketitle

\section{Introduction}
\label{sec.intro}
High thermal conductivity~($\kappa$) realized by polymers 
attracts attentions mainly because it 
can reduce the weight of mobile devices with 
electrically insulating properties, not like 
metallic materials.~\cite{2003MAS} 
Though the conductivities of polymers are usually 
lower than those of metals,\cite{2006WAN} 
they can be enhanced by making the ordering of 
the molecular's orientations (crystallinity). 
It has been reported for polyethylene~(PE) that 
the conductivity along the elongating direction 
of the molecules gets to be comparable to 
those of metals by increasing crystallinity.\cite{1985CHO} 
High crystallinity is, however, known to be 
difficult to be achieved experimentally~\cite{1993CHO}, 
leading lack of the contents of the database 
usable for the Materials Informatics ~(MI)
~\cite{2017IKE}
, searching for 
such polymers with high thermal conductivities. 
Theoretical estimations of such conductivities 
would assist to provide data over a lot of 
polymers to screen out the synthesis targets. 

\vspace{2mm}
Preceding theoretical works include 
several model analysis~\cite{1965HAN,1987FRE} 
as well as those by empirical molecular 
dynamics.~\cite{2008HEN,2009HEN}
The predictions were, however, found 
to be seriously depending on the choice of 
empirical force fields~\cite{2017NIN}. 
{\it Ab initio} approaches ~\cite{2016NAK, 2017NAK} are expected to 
exclude such ambiguity, and are getting 
to be feasible realized by the implementations 
of phonon analysis especially those beyond 
the harmonic approximation~\cite{phono3py}.
Such approaches have been applied to 
inorganic crystalline materials such as 
semiconductors,~\cite{phono3py,2009WAR,2015TAD2} 
achieving the predictions fairly coinciding 
with experiments over the range, 
1-10$^2$~W/mK.~\cite{phono3py}
For polymers, however, there has been little 
applications in spite of the industrial demands 
as described above, though there is 
a preceding work~\cite{2017NIN} 
applying the framework to evaluate 
the conductivity of PE as a typical prototype.

\vspace{2mm}
In this study, we evaluated the conductivity 
over such typical polymers for the 
exhaustive search for high $\kappa$, 
including PE, polyphenylene sulfide(PPS) and 
polyethylene terephthalate(PET). 
Those initial structures can be taken from 
the database of PolymerGenome 
project~\cite{2016HUA,2014SHA,2016MAN1}. 
We began with the calibrating discussions 
on PE, verifying our implementations 
to check quantitative consistency with 
available experimental $\kappa$ of 
crystalline and fiber PE.~\cite{1993CHO,1999CHO} 
Confirming the fairly well coincidence on PE, 
we applied the framework to a couple of polymers 
with typical structures being different from 
each other in terms of the number of benzene rings. 
Finding the PE structure (zero benzene ring) being 
the best to achieve high $\kappa$, 
we concentrated on the polymer structure 
without ring.
We took MI-like screening based on
the cheaper calculations of bulk modulus
to extract candidates for further evaluations of $\kappa$.
(The candidates are listed in Table~\ref{table.chosen}.)
We finally found that PVDF-beta polymer could 
achieve higher $\kappa$ than that of PE below around 80~K.

\section{Model and Methods}
\label{sec.det.spec}
Heat carriers to contribute to $\kappa$ include 
charged particles~\cite{BOOK1996QTOS, BOOK2001EAP},
diffusing atoms~\cite{1967GEO}, magnons~\cite{1982ANI}, and phonons~\cite{1982ANI, BOOK1996QTOS, BOOK2001EAP, BOOK1990TPOP}. 
The first three ingredients are excluded because 
the polymers considered here are insulating without any 
defects and non-magnetic. 
Only the phonons are to be formulated for $\kappa$ of 
polymers,~\cite{2006WAN} 
which is the Boltzmann equation applied to 
phonon flow.

\vspace{2mm}
The relaxation-time approximation
~\cite{BOOK1990TPOP, phono3py} 
of the Boltzmann equation leads to 
\begin{eqnarray}
  \kappa  = \frac{1}{{N{V_0}}}\sum\limits_\lambda
          {{C_\lambda }{{\bf{v}}_\lambda } \otimes }
          {{\bf{v}}_\lambda }\tau_\lambda ^{{\rm{SMRT}}}.
  \label{eq.1}
\end{eqnarray}
Here, $N$ is the number of unitcells used for the simulation
and $V_0$ is the volume of a unitcell.
$C_\lambda$, $\bm{v}_\lambda$, and  ${\tau}_\lambda$
are the specific heat at constant volume, group velocity,
and relaxation time of a phonon mode indexed by $\lambda$
(the upper index of $\tau_\lambda^{{\rm{SMRT}}}$ 
stands for '\underline{S}ingle-\underline{Mo}de 
\underline{R}elaxation-\underline{T}ime').

Relaxation time, $\tau$, appearing in Eq.(\ref{eq.1}) 
is given by the inverse of the imaginary part of 
the self-energy of the phonon, 
\begin{align}
  {\Gamma _\lambda }
  = &\frac{{18\pi }}{{{\hbar ^2}}}\sum\limits_{\lambda '\lambda ''}
  {{{\left| {{\Phi _{ - \lambda \lambda '\lambda ''}}} \right|}^2}}
  \nonumber\\
  &\left\{ {\left( {{n_{\lambda '}} + {n_{\lambda ''}} + 1} \right)
    \delta \left( {\omega- {\omega _{\lambda '}} -{\omega _{\lambda ''}}}\right)}
  \right.
  \nonumber\\
  &\left.{ + \left( {{n_{\lambda '}} - {n_{\lambda ''}}} \right)
    \left[ {\delta \left( {\omega  + {\omega _{\lambda '}}
          - {\omega _{\lambda ''}}} \right) - \delta
        \left( {\omega  - {\omega _{\lambda '}}
          + {\omega _{\lambda ''}}} \right)} \right]}\right\} \ , 
  \label{eq.2}
\end{align}
as ${\tau_\lambda } = 1/2{\Gamma_\lambda }\left( {{\omega _\lambda }} \right)$.
~\cite{wallace1998thermodynamics, 1962MAR}
The analytic form of ${\Gamma_\lambda}$ is given by 
perturbation theories with respect to the 
electron-phonon interactions~\cite{wallace1998thermodynamics,1962MAR}. 
To calculate ${\Gamma_\lambda}$ quantitatively, 
one requires to evaluate the lattice anharmonic contribution 
to the energy\cite{phono3py}, which is feasible 
by {\it ab initio} methods. 
The anharmonic contribution
~\cite{2017TRA}
is essential to get finite conductivity 
because the harmonic approximation leads to the superposition 
of phono motions, getting infinite conductivity without 
any collisions of phonons. 
A coefficient $b$ in an expansion of potential $U$
by an inter-atomic distance $u$:
\begin{eqnarray}
  U\left( {{u_0} + \tilde u} \right)
  = U\left( {{u_0}} \right) + a{\tilde u^2}
  + b{\tilde u^3} +  \cdots \  ,
  \label{eq.4}
\end{eqnarray}
is actually appearing in ${\Gamma_\lambda}$ 
because the term gives the scattering processes 
of phonon operators in the second quantization 
formalizm when one substitute the operator
representations of displacement, $u$.

\begin{figure}[htbp]
  \vspace{-30mm}
  \includegraphics[width=\hsize]{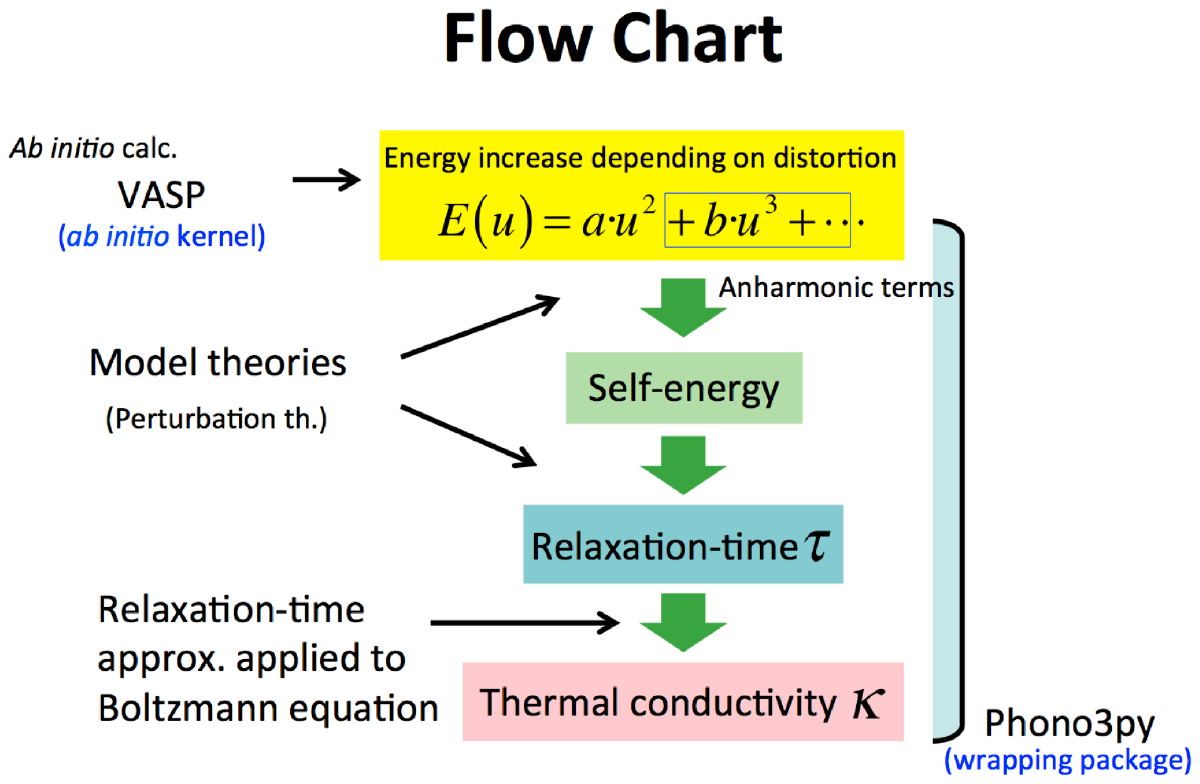}
  \vspace{-35mm}
  \caption{
    \label{flowChart}
    This is the workflow to evaluate thermal
    conductivity based on {\it ab initio} calculations.
    The harmonic and anharmonic coefficients
    $a,~b$ are obtained by {\it ab initio} calculations.
    The thermal conductivity is evaluated from
    the coefficients by solving a Boltzmann equation
    under relaxation-time approximation.\cite{phono3py}
  }
\end{figure}

\vspace{2mm}
The coefficients, $a,~b$ are obtained by density
functional theory (DFT):~We used VASP~\cite{1993KRE, 1996KRE, 1996KRE2, 1999KRE}
and employed projector augmented waves (PAW) method.
~\cite{1994BLO} We chose vdW-DF2 functional~\cite{2010LEE}
in order to describe van der Waals forces well.
We used phono3py package~\cite{phono3py} to set up
the calculations and compile up the {\it ab initio}
results to obtain specific heat capacity, phonon lifetime,
and thermal conductivity.
The schematic diagram to obtain heat capacity is described in Fig.~\ref{flowChart}. 

\section{Results and discussion}
\label{sec.results}
\subsection{Calibrations on Polyethylene}
\label{sec.results.a}
For PE, there are a lot of reference values of $\kappa$
available from both experiments
~\cite{1989PIR,1977CHO,1993CHO,1999CHO,2010SHE,1998FUJ,1975BUR,1972HAN,2005YAM,2017RON} 
and {ab initio} calculations.~\cite{2017NIN, 2012JIN}
Thus, we will check the reliability of our framework
for the system as an example.
The references include both those of 'fibers' 
and 'crystals' (the reason why we put apostrophe 
on it is explained below): 
A polymer's crystal means the ordered stacking of 
1-dim. fibers, that is quite difficult to be 
synthesized experimentally. 
Practical samples are actually the mixtures 
including disordered 1-dim. fibers as well as 
fragments of crystals in a polycrystalline 
manner. 
The experimental reference values of '1-dim. fibers' mean 
the extrapolation of the observed 
values towards the limit by increasing the 
order of orientations of fibers. 
For PE, such extrapolations are 
performed by controlling the volume fraction.~\cite{1993CHO}
Values of 'crystals' are obtained by some model 
formulae such as that by Halpin-Tsai.\cite{1976HAL}
By putting values observed for the mixture of fibers 
as well those of amorphous, 
the formulae give the estimations for crystals. 
~\cite{1969HAL, 1974NIE}

\begin{figure}[htbp]
  \vspace{-30mm}
  \includegraphics[width=\hsize]{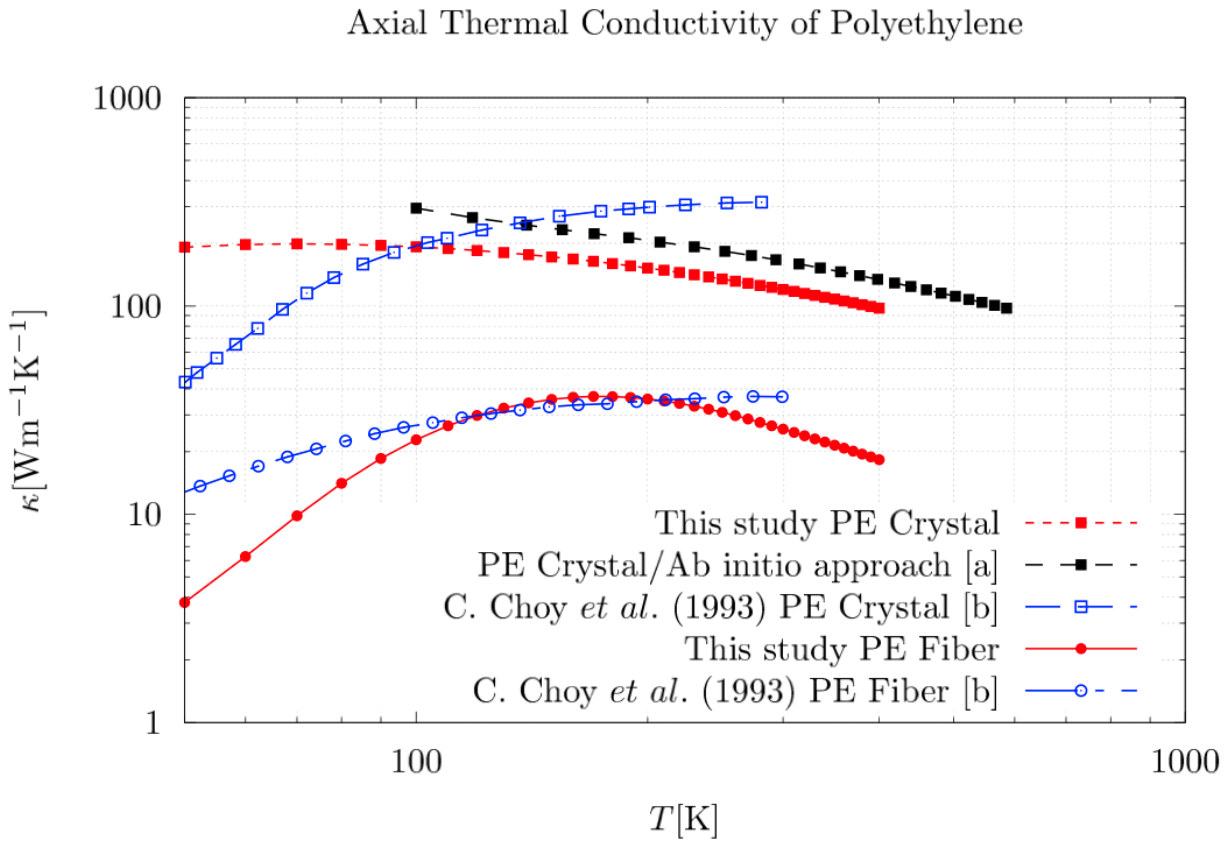}
  \vspace{-35mm}
  \caption{[pe\_chain]
  (a:\cite{2017NIN}, b:\cite{1993CHO})
    Thermal conductivities of the polyethylene in the 3-dim. crystal 
and the 1-dim. chain, as a function of the temperature, 
compared with experiments and other preceding works.
The present study fairly reproduces the quantitative 
difference by an order between those  
in crystals and fibres observed experimentally. 
Though the temperature dependence differs from 
the experimental 'crystal' case significantly, 
it fairly coincides with the preceding ab initio study. 
We note that the experimental data provided as 'crystal' 
is not taken from true crystals, but estimated by extrapolations 
from the data of fibre data. 
  }
  \label{pe_chain}
\end{figure}

\vspace{2mm}
Our estimation of the specific heat at 
constant volume of PE gets 
$C_v$~=~1.653 [Jg$^{-1}$K$^{-1}$] at $T=310.15~[K]$ 
obtained by phonopy package~\cite{phonopy}, 
being in fairly agreement with 
the experimental value at the same temperature, 
1.479 $[Jg^{-1}K^{-1}]$~\cite{polyInfo1}.
The coincidence would ensure that 
our framework works quantitatively well 
at least upto the extent of the harmonic 
approximation. 
Fig.~\ref{pe_chain} shows the comparisons 
of thermal conductivities ($\kappa (T)$) of our estimations 
and preceding theoretical/experimental values 
for 'crystals' and fibers. 
Quantitative coincidence seem fairly well, 
especially reproducing the difference 
by an order between crystal and fiber. 
Though that is consistent within the 
theoretical works between ours and 
preceding works~\cite{2017NIN}, 
the temperature dependences seem to behave 
in quite different way when we compare 
with experimental ones. 
To discuss the difference, 
we would like to concentrate only on 
fiber cases. 
For 'crystals', the analysis gets to be 
too complicated because the consequences 
come from behind the empirical model formulae 
as we explained above, inherently 
including many ambiguities (sample qualities, 
model approximations {\it etc}). 
Furthermore on crystals, 
the consistency within theoretical studies  
would support the justification of 
the present work. 

\begin{figure}[htbp]
  \vspace{-30mm}
  \includegraphics[width=\hsize]{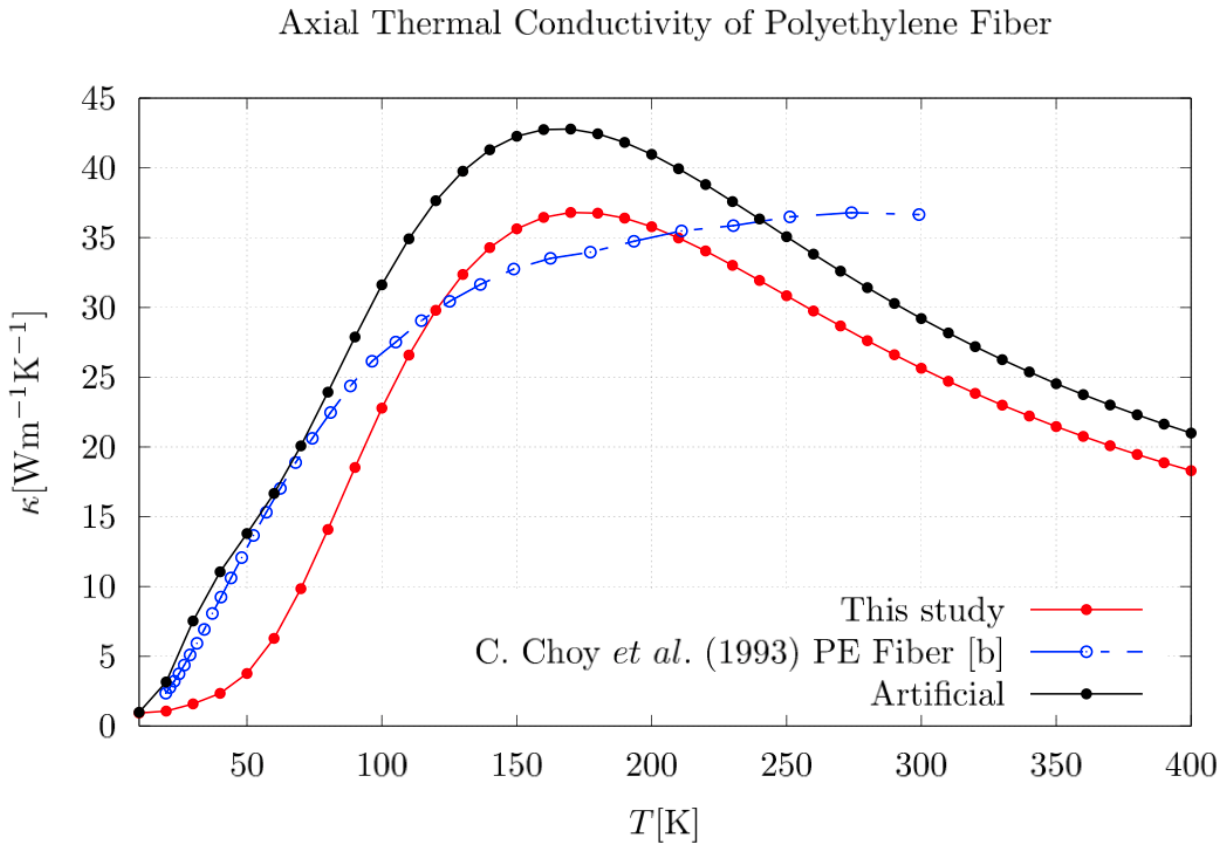}
  \vspace{-35mm}
  \caption{[fig.results.pe\_chain.pe\_chain.1]
      (b:\cite{1993CHO})
    Temperature dependences of the thermal conductivity, $\kappa (T)$, 
    of polyethylene fibers. The blue plot represents
    experimental data, while the red does our theoretical
    results for an isolated 1-dim. fiber. 
    The black plot corresponds to the case with the artificially 
    modified distribution of the mode contributions (see text).    
  }
  \label{fig.results.pe_chain.pe_chain.1}
\end{figure}

\begin{figure}[htbp]
  \vspace{-30mm}
  \includegraphics[width=\hsize]{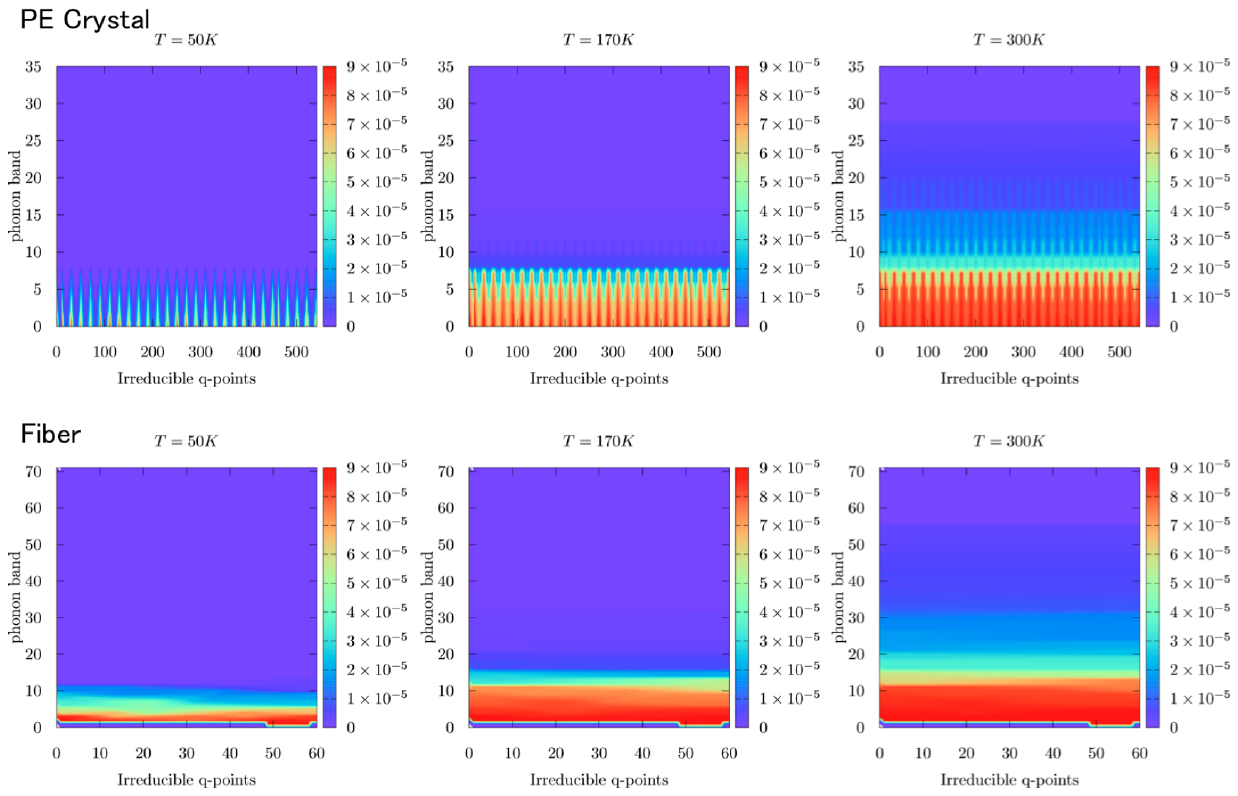}
  \vspace{-35mm}
  \caption{
    These graphs show $C_v$ of each phonon mode
    designated by indices of $q$-point and ones of phonon band,
    for PE 3-dim. crystal (upper) and 1-dim. fiber (lower) under different
    temperatures: $T$=50,~170,~300~K.
  }
  \label{fig.results.Cv_PE_c_and_f}
\end{figure}

\begin{figure}[htbp]
  \vspace{-30mm}
  \includegraphics[width=\hsize]{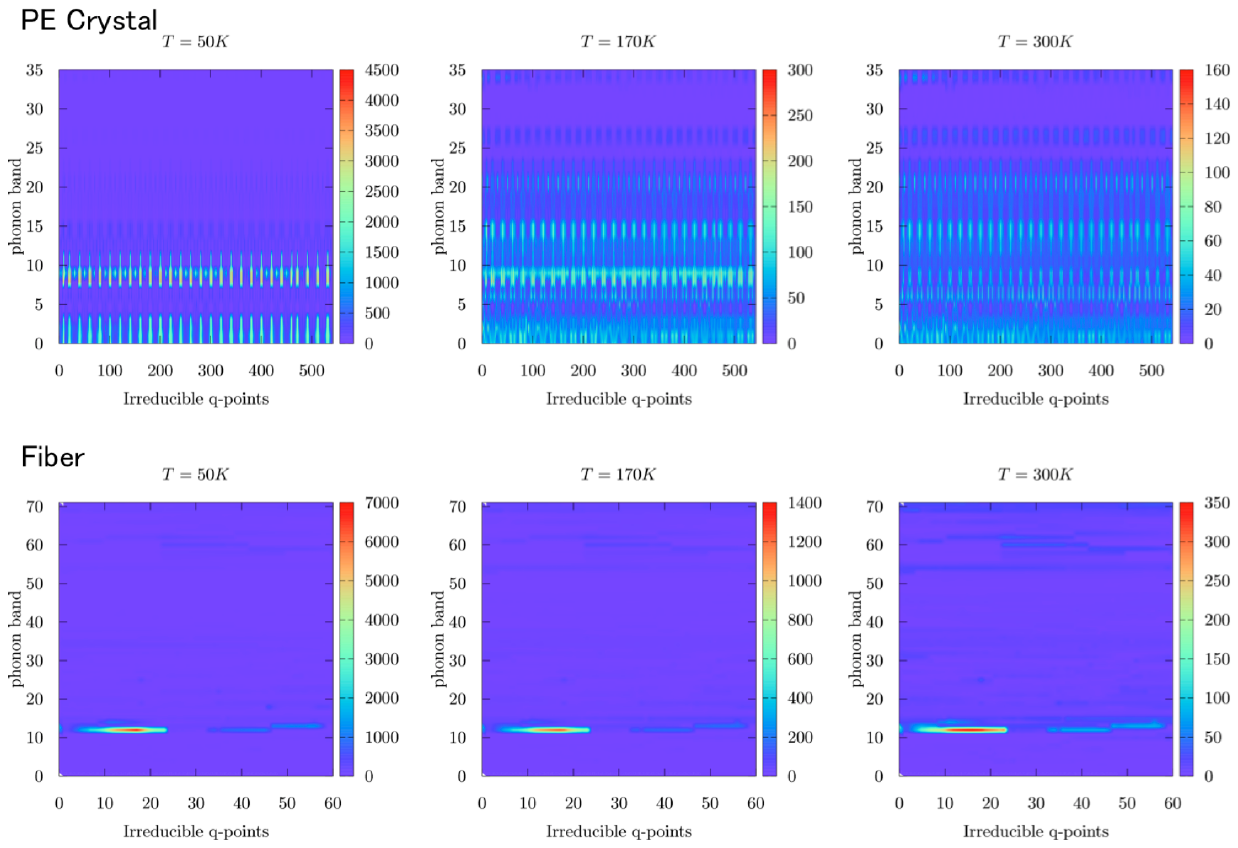}
  \vspace{-35mm}
  \caption{
    [fig.results.tau\_PE\_c\_and\_f]
    These graphs show relaxation-time of each phonon mode
    designated by indices of $q$-point and ones of phonon band,
    for PE 3-dim. crystal (upper) and 1-dim. fiber (lower) under different
    temperatures: $T$=50,~170,~300~K.
    \label{fig.results.tau_PE_c_and_f}
  }
\end{figure}

\vspace{2mm}
Fig.~\ref{fig.results.pe_chain.pe_chain.1} 
shows the detailed comparison of the 
temperature dependence, $\kappa (T)$, 
between our simulation~(red) and the 
experiment~(blue)\cite{1993CHO}. 
The difference between experiments and 
the simulation would be attributed to 
the dimensionality. 
While the simulation treats an 
isolated 1-dim. fiber, 
the practical experiments 
treat the bunch of fibers 
which includes inter-fiber contributions. 
Since the contribution is also 
included in the crystal case, 
the comparison between the fiber and the crystal 
within our simulation would help 
to understand why the $T$-dependence 
between red and blue plots in 
Fig.~\ref{fig.results.pe_chain.pe_chain.1} 
differs each other. 
In Eq.~(\ref{eq.1}), we can decompose 
$\kappa (T)$ into the mode contributions 
under the summation over the mode index, 
$\lambda =(b,q)$, where $b$ and $q$ denote 
the indices for band and wave vector, respectively. 
Since in the simulation there is no $T$-dependence 
of $\vec v_{\lambda}$ (group velocity), 
the dependence of $\kappa (T)$ is determined by 
those of $C(T)$ and $\tau (T)$. 
Fig.~\ref{fig.results.tau_PE_c_and_f} and 
\ref{fig.results.Cv_PE_c_and_f} 
show the distributions of the mode 
contributions for the relaxation time 
and the specific heat, respectively, 
shown as the intensity 
on the $(b,q)$-plane. 
The brighter region corresponds to 
the 'active range', where 
the larger contributions distribute. 
We can find in Fig.~\ref{fig.results.tau_PE_c_and_f} 
that the active range gets to be largely 
restricted for the fiber case when compared to 
crystal case. 
This difference would be attributed to 
the difference in the $T$-dependences on 
Fig.~\ref{fig.results.pe_chain.pe_chain.1}: 
When we force the modes activated by hand 
artificially so that their distributions 
for fibers can get to the same as those of 
crystal, the $T$-dependence gets to be 
closer to the experimental case in lower 
temperature range, as shown by the black 
plots in Fig.~\ref{fig.results.tau_PE_c_and_f}. 
This would support that the $T$-dependence 
is dominated by the dimensionality, since
the difference of the distributions for fiber
and crystal cases reflects their difference
in the dimensionality, namely whether the inter-fiber
transportation exists or not.

\vspace{2mm}
It is interesting that $\kappa(T)$ of PE fiber looks
changing in proportion to $\sim T^3$, although
the sample size effect~\cite{1999CHO, kittel_ssp}
does not exist in the simulation.
This fact indicates that the change ratio of $\tau(T)$
for temperature is much smaller than that of $C(T)$.
This apparently contradicts that phonon life time
diverges towards infinity when temperature decreasing
in general. However, our system is a fiber and
the phonons could be scattered on the wall of fiber.
This could restrict the divergence of $\tau(T)$ and
let $\kappa(T)$ follow $\sim T^3$.

\vspace{2mm}
The magnitude relationship between 
experiments~(blue) and simulations~(red) 
in Fig.~\ref{fig.results.pe_chain.pe_chain.1} 
turns from 'exp.$>$~sim.' to 'exp.$<$~sim.', 
and again to 'exp.$>$~sim.' as $T$ increases. 
'The more active range of $\tau (T)$ 
for crystal (a model for experiments 
with larger dimensionality)', as explained above, 
would be the plausible reason at the lower $T$ range. 
In the 1-dim. fiber case, the modes within 
the narrow active range have remarkably 
large relaxation time while in the crystal case 
the wider range consists of those with 
moderately large relaxation time, 
just like a 'conservation rule', 
as we found by carefully inspections of 
$\tau$ for both cases. 
$C(b,q;T)$, shown in Fig.~\ref{fig.results.Cv_PE_c_and_f}, 
plays as the 'mask' put over the distribution of 
$\tau (b,q;T)$, by forming a product to give $\kappa$. 
The mask gets to have the weight around 
the active range of $\tau$ by increasing $C(T)$ 
at the mode around there, especially 
around $T\sim$ 170~K. 
This leads to the overtaking to get 'exp.$<$~sim.' 
at the middle range of $T$. 
Generally speaking, $\kappa (T)$ should 
decay at the high temperature range
~\cite{1999CHO,kittel_ssp}
due to the increasing of the phonon scattering.
Simulation results (red) surely follows
this consequence but the experimental ones (blue)
does not meanwhile, which leads to 'exp.$>$~sim' again.
The difference would explained by the dimensionality:
There are more number of phonon paths in the experimental
situation than in the simulation, owing to the thermal
transport among fibers. The extra paths could help for
phonons to avoid colliding each other and suppress
phonon scattering.

\subsection{Screening for high $\kappa$ polymers}
\label{sec.results.b}
Fig.~\ref{fig.kappa} compares the calculated $\kappa(T)$
for the chosen polymers. The filled points corresponds
to the results for the typical polymers, PE, PPS, and PET,
where PE has much higher $\kappa(T)$ than the others.
This would be because only PE does not include benzene rings,
which is why we intensively surveyed $\kappa(T)$ of the six
linear polymers.
Among them, PVDF-beta and PVFP has comparatively high thermal
conductivity.
It is further surprising that the $\kappa(T)$ of PVDF-beta
is superior to that of PE below $\sim$80~K.
It is explained from the difference of the mappings of
phonon relaxation time:
Fig.~\ref{fig.0025_tau_3d_50Kd128} represents that
the phonon modes having long relaxation time
are concentrated in the low eigen energy region
in the case of PVDF-beta crystal at 50 K. 
Conversely, the left upper figure of
fig.~\ref{fig.results.tau_PE_c_and_f},
represents that such modes are distributed in a wide
range of eigen energy in the case of PE crystal.
Under low temperature, only phonon modes with low
eigen energy have high $C_v$ so PVDF-beta crystal
has higher thermal conductivity. On the other hand,
when the temperature increases, the $C_v$ of phonon
modes with middle to high eigen energy get increased
and PE crystal has higher thermal conductivity eventually.

\begin{figure}[htbp]
  \vspace{-30mm}
  \includegraphics[width=\hsize]{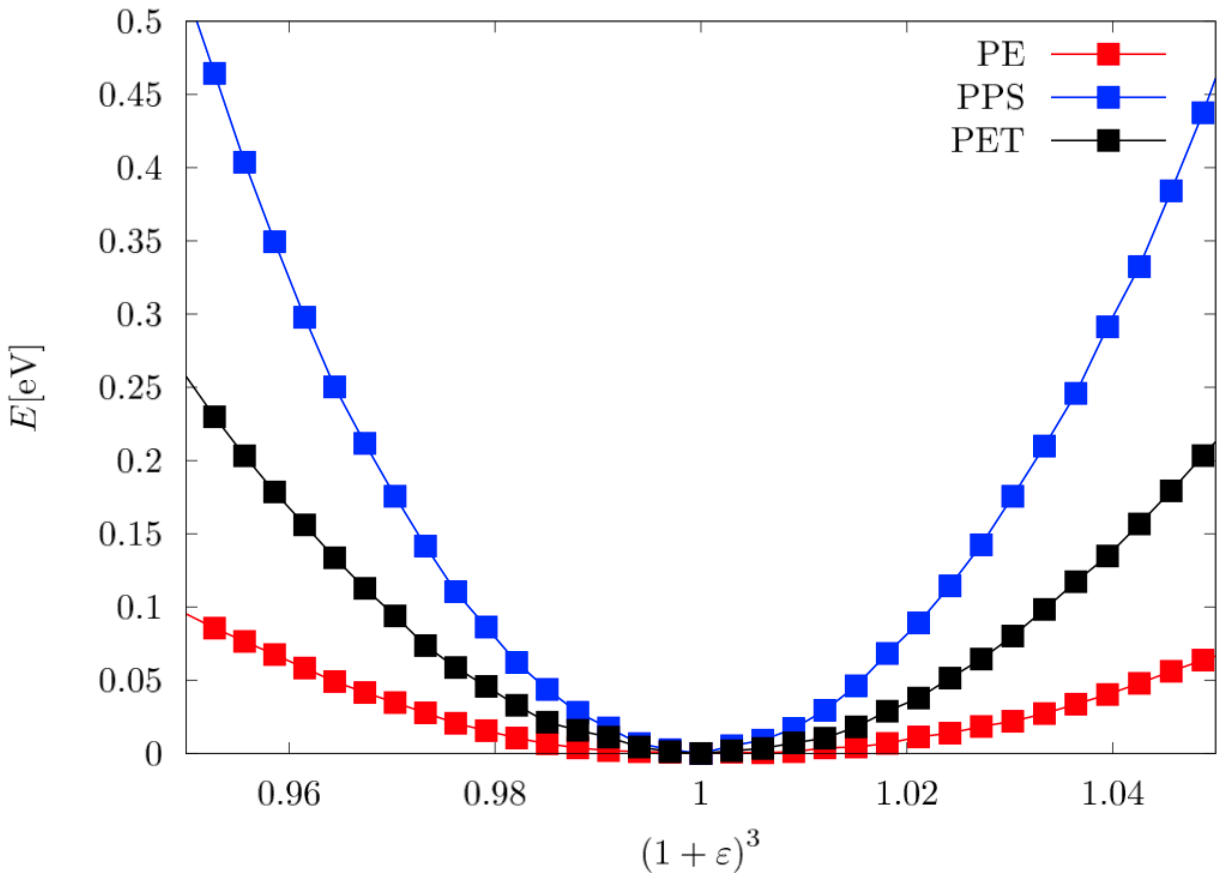}
  \vspace{-35mm}
  \caption{
    The dependence of the ground state energies 
    on the volume, $E(V)$. 
    The plots are given as a function of $\varepsilon$, 
    where $V=(1+\varepsilon)^3V_0$ with the equilibrium 
    volume, $V_0$. 
  }
  \label{fig.results.pe_pps_pet}
\end{figure}

\begin{figure}[htbp]
  \includegraphics[width=\hsize]{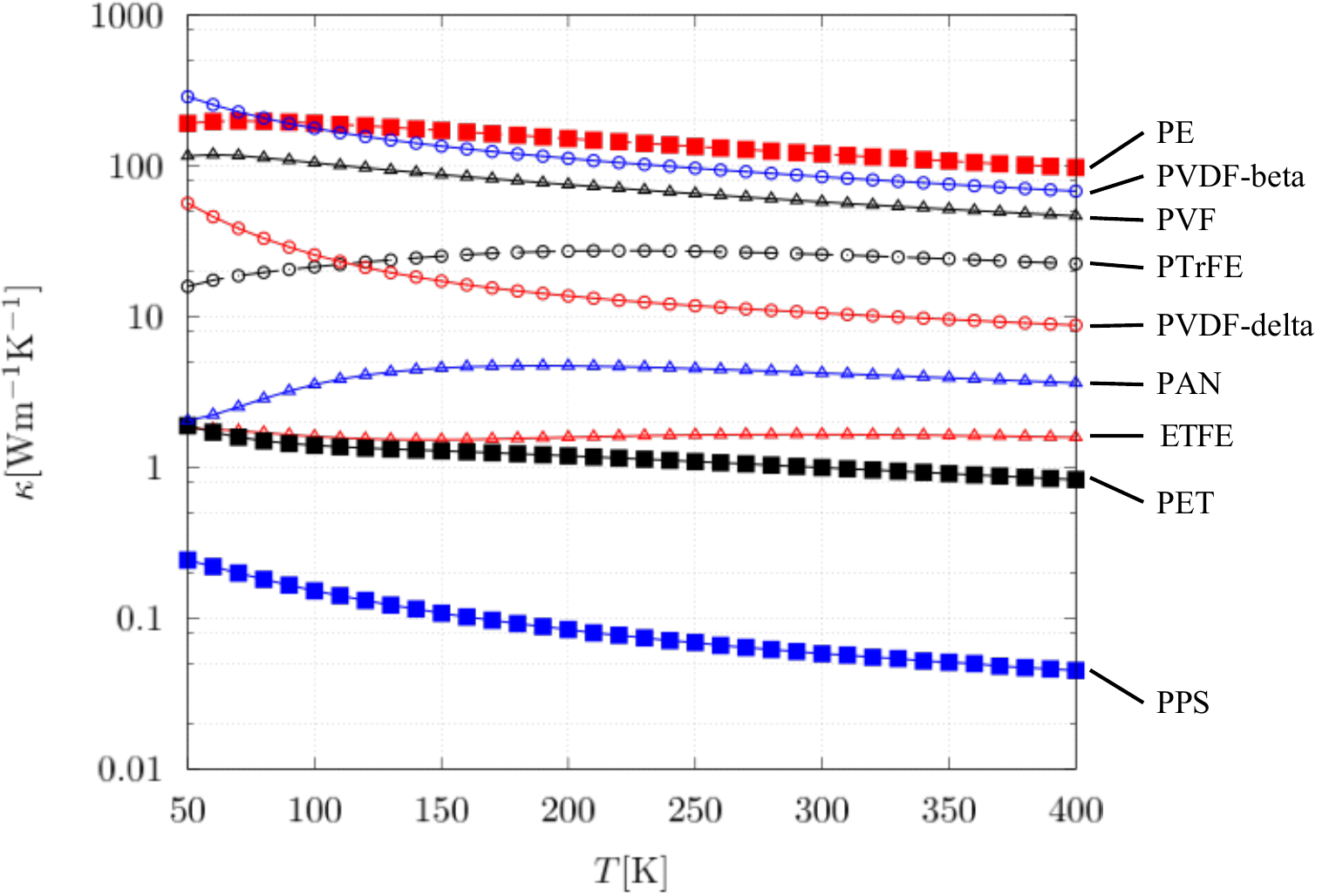}
  \caption{[fig.kappa]
    This figure compares crystalline thermal
    conductivity of the selected polymers.
    The linear polymers without benzene rings
    tend to have higher thermal conductivity
    than PET and PPS.
    PVDF-beta and PVF have impressively high
    thermal conductivity comparable with PE.    
  }
  \label{fig.kappa}
\end{figure}

\vspace{2mm}
Though we performed eventually the evaluations of $\kappa$ 
for all the cases, it is quite expensive to 
perform it for the entire possible structures 
of polymers in general. 
It is therefore desirable to establish 
the way of the screening for the higher $\kappa$ 
with cheaper computational costs. 
Since for $\kappa$ the anharmonicity matters, 
the trend of the magnitude of it would be 
captured by the Gruneisen parameter, 
being a typical measure for the anharmonicity.
~\cite{ziman_pottos} 
The parameter relates to the bulk modulus,~\cite{1994VOC}
and hence the trend is expected to be captured 
by the quantity, for which the required computational cost 
gets to be saved a lot. 
Fig.~\ref{fig.results.pe_pps_pet} corresponds to 
the estimations of the bulk modulus, and 
we clearly see that the trend of the modulus, 
namely the curvatures of $E(V)$ well correlates 
with that of $\kappa$ given in 
Fig.~\ref{fig.kappa} 
(the lower the modulus, the higher the $\kappa$, 
as expected from the relation~\cite{1994VOC}).

\begin{figure}[htbp]
  \vspace{-30mm}
  \includegraphics[width=\hsize]{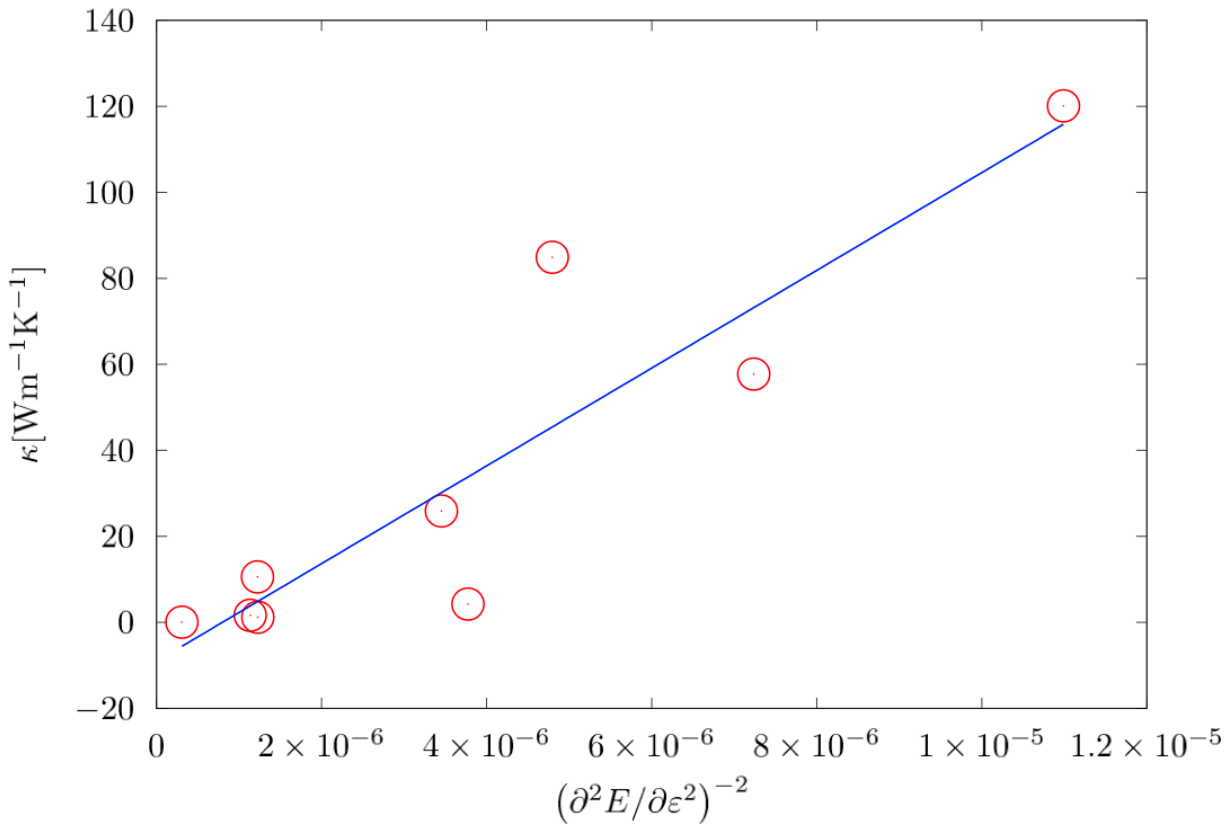}
  \vspace{-35mm}
  \caption{[fig.corr]
    This figure maps thermal conductivity and bulk modulus 
    ${\left( {\partial^2 E/\partial \varepsilon^2 } \right)^{-2}}$ 
    for crystals of the selected polymers
    ($\varepsilon$ is defined in Fig.~\ref{fig.results.pe_pps_pet}).
    A linear relationship is observed clearly
    between both quantities, where the fitting
    line is given by least-square method and
    its formula is $y = 1.137 \times 10^{7} x - 9.075$.
  }
  \label{fig.corr}
\end{figure}

\begin{figure}[H]
  \vspace{-20mm}
  \includegraphics[width=\hsize]{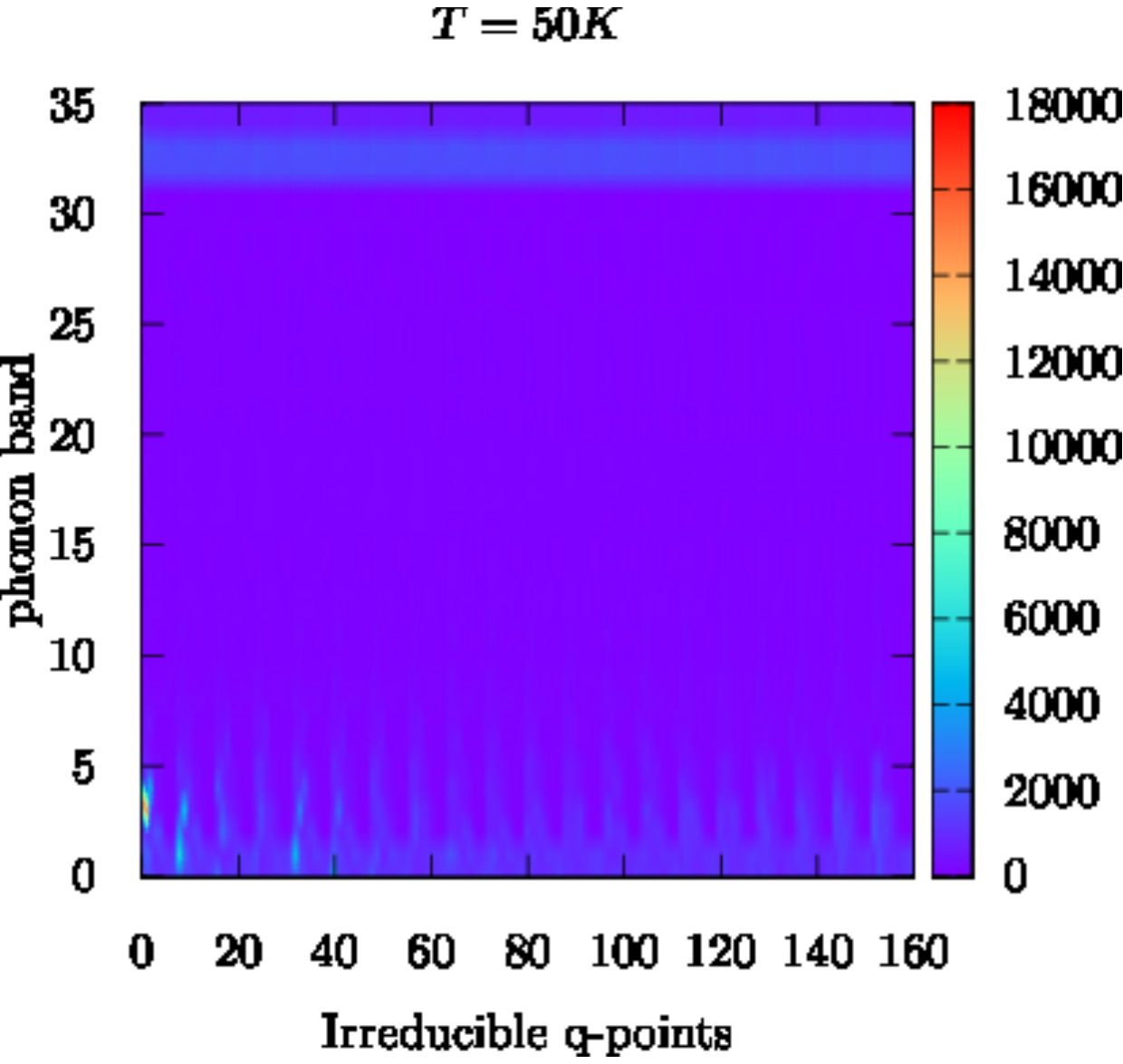}
  \vspace{-25mm}
  \caption{
    [fig.0025\_tau\_3d\_50Kd128]
    This graph maps the relaxation time of each phonon mode
    for PVDF-beta crystal at 50~K.
    The horizontal axis represents the indices of $q$-point
    and the vertical axis does those of phonon band.
  }
  \label{fig.0025_tau_3d_50Kd128}
\end{figure}

Fig.~\ref{fig.corr} proves that the modulus (curvature) 
would be a good descriptor for the $\kappa$ for 
the polymers, achieving the correlation coefficient 
$\sim$0.9045.
\begin{table}[htbp]
\begin{center}
\caption{[table.chosen] Our target polymers are listed in this table with their abbreviations.}
\label{table.chosen}
\begin{tabular}{ccc} \hline
Name & Abbreviation \\ \hline
Poly(ethylene)&PE\\
Poly(phenylene sulfide)&PPS\\
Poly(ethylene) terephthalate&PET\\
Poly(vinylidene fluoride)-delta&PVDF-delta\\
Poly(vinylidenesurely fluoride)-beta&PVDF-beta\\
Poly-trifluoroethylene&PTrFE\\
Poly(tetrafluoroethylene-alt-ethylene)&ETFE\\
Polyacrylonitrile&PAN\\
Poly(vinyl fluoride)&PVF\\
\hline
\end{tabular}
\end{center}
\end{table}


\section{Conclusion}
\label{sec.conc}
Thermal conductivity of polymer crystals have almost 
not been studied because of the difficulty of synthesis~\cite{1993CHO, 1999CHO}. 
We investigated thermal conductivity for some polymer 
crystals by first principles calculations~\cite{phono3py}.
Our calculation successfully reproduced that PE crystal 
has higher thermal conductivity by one digit than its fiber~\cite{1993CHO}.
On the other hand, the predicted temperature dependence of 
thermal conductivity of fibers is qualitatively different 
from the experimental ones~\cite{1993CHO}. 
This is because the thermal conduction among fibers are 
not excluded in the experiments unlike the simulations.
According to the ab initio predictions for PE, PPS, and PET crystals,
it is established that PE crystal has much higher thermal conductivity
than the others.
Therefore we surveyed thermal conductivities of the six polymer crystals
similar to PE in addition, and we revealed that PVDF-beta and PVF crystals
have comparatively high thermal conductivities with PE crystal.
We have also found out an interesting law:
The predicted thermal conductivities are clearly in proportion to 
the curvature of energy-volume curve.
Since it is much cheaper to predict the latter value, 
this law would be useful to discovery a new polymer crystal
having further high thermal conductivity.@

\section{Acknowledgments}
The computation in this work has been performed 
using the facilities of the Research Center for Advanced Computing Infrastructure (RCACI) at JAIST.
T.I. is grateful for financial suport from Grant-in-Aid for JSPS Research Fellow (18J12653).
K.H. is grateful for financial support from a KAKENHI grant (JP17K17762), 
a Grant-in-Aid for Scientific Research on Innovative Areas ``Mixed Anion'' project (JP16H06439) from MEXT, 
PRESTO (JPMJPR16NA) and the Materials research by Information Integration Initiative (MI$^2$I) project 
of the Support Program for Starting Up Innovation Hub from Japan Science and Technology Agency (JST). 
R.M. is grateful for financial supports from MEXT-KAKENHI (17H05478 and 16KK0097), 
from Toyota Motor Corporation, from I-O DATA Foundation, 
and from the Air Force Office of Scientific Research (AFOSR-AOARD/FA2386-17-1-4049).
R.M. and K.H. are also grateful to financial supports from MEXT-FLAGSHIP2020 (hp170269, hp170220).


\bibliographystyle{apsrev4-1}
\bibliography{references}
\end{document}